\documentclass[prb,twocolumn,showpacs,preprintnumbers,amsmath,amssymb]{revtex4}

\usepackage{graphicx}
\usepackage{dcolumn}
\usepackage{bm}
\usepackage{epsfig}
\usepackage{subfigure}

\begin{document}


\title{Cobalt nano-clusters on metal supported Xe monolayers: influence of the substrate on cluster formation kinetics and magnetism}

\author{V. Sessi}
\email{v.sessi@fkf.mpg.de}
\affiliation{Max-Planck-Institut f\"ur
Festk\"orperforschung, Heisenbergstrasse 1, 70569 Stuttgart,
Germany}
\author{K. Kuhnke}
\affiliation{Max-Planck-Institut f\"ur Festk\"orperforschung,
Heisenbergstrasse 1, 70569 Stuttgart, Germany}
\author{A. Enders}
\affiliation{Dept.~of Physics and Astronomy and MCMN, University of Nebraska, Lincoln, NE 68588, USA}
\author{J.~Zhang}
\affiliation{Max-Planck-Institut f\"ur Festk\"orperforschung, Heisenbergstrasse 1, 70569 Stuttgart, Germany} \affiliation{School of
Material Science and Engeneering, Hebei University of Technology, Tianjin 300130, China}
\author{P.~Bencok}
\affiliation{Diamond Light Source Ltd, Diamond House, Harwell
Science and Innovation Campus, Didcot, Oxfordshire, OX11 0DE, Great
Britain}
\author{S. Bornemann}
\affiliation{Department Chemie und Biochemie,
Ludwig-Maximilians-Universit\"at M\"unchen, 81377 M\"unchen,
Germany}
\author{J. Min\'{a}r}
\affiliation{Department Chemie und Biochemie, Ludwig-Maximilians-Universit\"at M\"unchen, 81377 M\"unchen, Germany}
\author{H. Ebert}
\affiliation{Department Chemie und Biochemie,
Ludwig-Maximilians-Universit\"at M\"unchen, 81377 M\"unchen,
Germany}
\author{J. Honolka}
\affiliation{Max-Planck-Institut f\"ur Festk\"orperforschung,
Heisenbergstrasse 1, 70569 Stuttgart, Germany}
\author{K. Kern}
\affiliation{Max-Planck-Institut f\"ur Festk\"orperforschung,
Heisenbergstrasse 1, 70569 Stuttgart, Germany}

\date{\today}

\begin{abstract}
The growth dynamics of submonolayer coverages of Cobalt during
buffer layer assisted growth on Ag(111) and Pt(111) substrates is
investigated by variable temperature scanning tunneling microscopy
in the temperature range between 80 and 150 Kelvin. It is found that
attractive cluster-substrate interactions can govern the cluster
formation on the Xe buffer layer, if the Xe layer is sufficiently
thin. The interpretation of the microscopy results are supported by
x-ray magnetic circular dichroism which monitors the effect of
cluster-substrate interactions on the formation of magnetic moments
and magnetic anisotropy of Co nanocluster during the different
stages of growth. {\it Ab-initio} calculations show that the cluster
magnetism is controlled by the interface anisotropy, leading to
perpendicular magnetization for Co on Pt(111). Limits of and new
potential for nanocluster fabrication by buffer layer assisted
growth are discussed.
\end{abstract}


\maketitle

\section{Introduction}
A fundamental physical understanding of magnetic nanostructures in contact with a medium is essential from a technologic point of
view: the magneto-crystalline anisotropy energy (MAE) and the superparamagnetic blocking temperature $T_{\text{B}}$ come to depend
not only on the nanostructure material itself but also on the electronic and magnetic properties of the supporting medium and in
particular on the contact area between the two. There is therefore a critical need for versatile synthesis strategies that permit
the fabrication of nanoclusters at surfaces with well defined magnetic functionality and with unprecedented control over the
cluster size, shape, size distribution, areal density and even positional accuracy.

The synthesis of nanoclusters by buffer layer assisted growth (BLAG)~\cite{WeaverScience} has become an increasingly attractive
strategy to form clusters at surfaces, both for its simplicity and for its potential to be a generic approach to fabricate clusters
of any material, also ordered cluster arrays, on all kinds of surfaces. Here, a noble gas layer acts as a buffer between the
substrate and deposited single atoms which self-assemble on the layer and form nanostructures. The role of the noble gas layer is
to decouple the cluster material from the substrate, so that the cluster formation is unimpeded by the substrate. It has been shown
in several publications that the final cluster size and the areal distribution only depend on the thickness of the Xe buffer layer
and the metal coverage, thus offering a desirable way to grow 'anything on anything' \cite{WeaverSurfSci,WeaverPRLdens.size}.
\newline \indent Recent experimental and theoretical work has shown, however, that also BLAG is crucially dependent on many
experimental parameter and that it is not a 'one size fits all' method to produce clusters. For instance, it has been shown that
the van der Waals interaction between the clusters and the substrate can effectively pull the clusters into the Xe layer in some
cases~\cite{WeaverPRBXPS}. It was also claimed that at least 60~ML of buffer layer is necessary to electronically decouple the
metal clusters from the substrate~\cite{Hövel}. Finally, wetting of the nanoclusters by Xenon and therefore partial embedding into
the Xenon layer has been predicted by molecular dynamics simulations~\cite{Marchenko}. These results rather indicate that the
cluster formation is the result of a yet to be understood, complex interplay between the atoms from the substrate, the buffer layer
and the deposited atoms.\newline \indent Nano-scale effects on the Co growth due to interactions with the surrounding material are
expected to be of paramount importance when the system is reduced to a sub-monolayer amount of metal on a very thin rare gas layer.
We have already demonstrated that indeed, under these conditions, the morphology of the substrate underneath can be used as a
template to tailor the cluster size and their lateral arrangement on the substrate~\cite{ZhangPt997 and hBN}.\newline \indent In
the present paper we investigate the mechanisms driving the dynamics during BLAG in detail, focusing on the early stage of cluster
formation. The combined use of variable temperature scanning tunneling microscopy (VT-STM) and x-ray magnetic circular dichroism
(XMCD) allows us to monitor directly the growth mechanism of Co/Xe in case of absorption on two electronically very different metal
substrates: Ag(111) and Pt(111). For a detailed description of the sample preparation and the experimental settings we refer the
reader to the appendix in section~\ref{Appendix: Experimental}. This work shows how the interface properties determine not only the
morphology but also the magnetic properties of the cluster system. The measured magnetic moments and magnetic anisotropy are
compared to {\it ab-initio} magnetic calculations of monolayer and bi-layer Co epitaxial island systems of various size deposited
on Ag(111) and on Pt(111). Among other things from this comparison it becomes evident that the MAE, that emerges upon contact with
the surface, is a pure interface effect.

\section{Cluster growth on thick Xenon buffer layers}
\label{Cluster growth on thick Xe buffer layers}

\subsection{Cluster growth and morphology}
\label{Cluster growth and morphology} 
\begin{figure}[htbp]
\includegraphics[width=.45\textwidth]{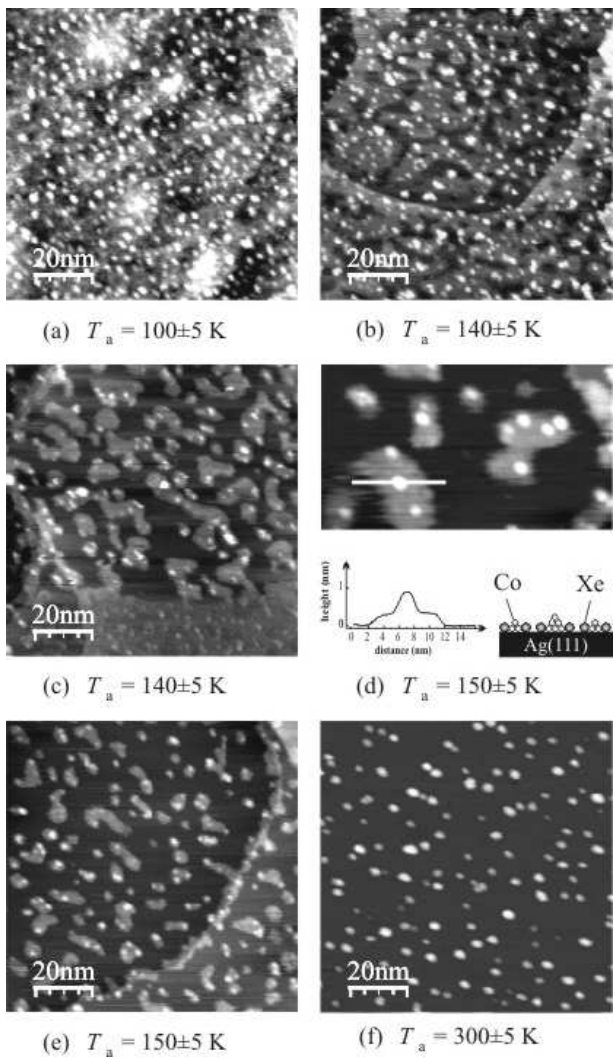}
  \caption{\label{VT-STM1} STM topographies for thick Xenon buffer
layers: Co/Xenon/Ag(111) deposited at 30K and then
annealed to progressively higher temperatures $T_{a}$ as indicated
in the caption;  (d) zoom-in of image (e), together with a line scan
over a Co cluster and a cartoon showing the Co immersed in Xe.}
\end{figure}

The morphology of Co clusters formed with comparatively thick Xe layers on Ag(111) was
studied with VT-STM. The Xe thickness was controlled by the exposure of the clean Ag(111) substrates to Xe partial pressures in UHV
at substrate temperatures of 30K. For determination of the Xe thickness we used an experimentally established estimate, 1 ML Xe =
5.5 Langmuir, from Ref.~\cite{Langmuir}(1~Langmuir = 1sec~$\cdot10^{-6}$~Torr). Thus, exposure of the substrate to 50~Langmuir (L)
resulted in Xe buffer layers of approximately 9 monolayers thickness. Cobalt was deposited on the Xe buffer layer at $T=30$K by
thermal evaporation from a Co rod. For the samples in this section 5\% of a full epitaxial Co monolayer have been
deposited.\newline \indent STM images were taken at different temperatures while warming up the sample to room temperature. In
Fig.~\ref{VT-STM1}, STM topography images are displayed taken at temperatures of (a) 100K, (b, c) 140K, (d, e) 150K, and (f) at
300K, after full Xe desorption. The images show clearly the desorption of the Xe buffer layer, and the presence and ripening of Co
clusters. In (a), small clusters of about 1-2nm diameter can already be resolved, on a rather noisy background. The overall quality
of the images is reduced by the diminished electron tunneling through the noble gas as well as the weak bonds of the adlayer atoms
to the substrate. However, it can clearly be seen how the Xe layer breaks up into islands surrounding the Co clusters at about 140K
(Fig.~\ref{VT-STM1}(b) and (c)). The transition from continuous to interrupted Xe layer can best be seen at the bottom of
Fig.~\ref{VT-STM1}(c). Several hours of time had elapsed between the acquisition of images (b) and (e). \newline \indent From the
images (a-e) we conclude that the desorption of Xe from Ag(111) occurs for temperatures above $T = 100$K and it is especially
pronounced in the temperature window between 140K and 150K. A delay of desorption of Xe is visible at defects such as substrate
steps and Co clusters. The Xe is completely removed at substrate temperatures above 150K. Thus, the observed desorption temperature
is significantly higher than that of bulk Xe, which is about 55K. It is well-known that atomically thin Xenon layers on metal
surfaces have increased desorption temperatures compared to bulk Xe, due to interactions with the substrate. Depending on initial
coverage, the Xe layer in direct contact with Ag(111) has been reported to desorb between 85K and 90K, while the top layer of a Xe
bi-layer starts desorbing at 62K~\cite{WandeltXPS}. The desorption process is of zeroth order, that is, it does not change its
phase during annealing.\newline \indent Tip convolution effects make it rather difficult to derive quantitative values for the size
of the clusters from topography data. Nevertheless, we estimate the temperature dependent cluster size knowing the number of Co
atoms on the surface from the nominal Co coverage of 0.05ML and the cluster density from STM images. We obtain an increasing
average number of atoms per cluster with temperature of about $20\pm 5$ atoms at 100K, $(40\pm 7)$ atoms at 140K and about $50\pm5$
atoms at room temperature. The given error here is the statistic error found after
 averaging over several topographies.

\subsection{Magnetism of Co clusters on Ag(111)}

\begin{figure*}[htbp]
\includegraphics[width=0.7\textwidth]{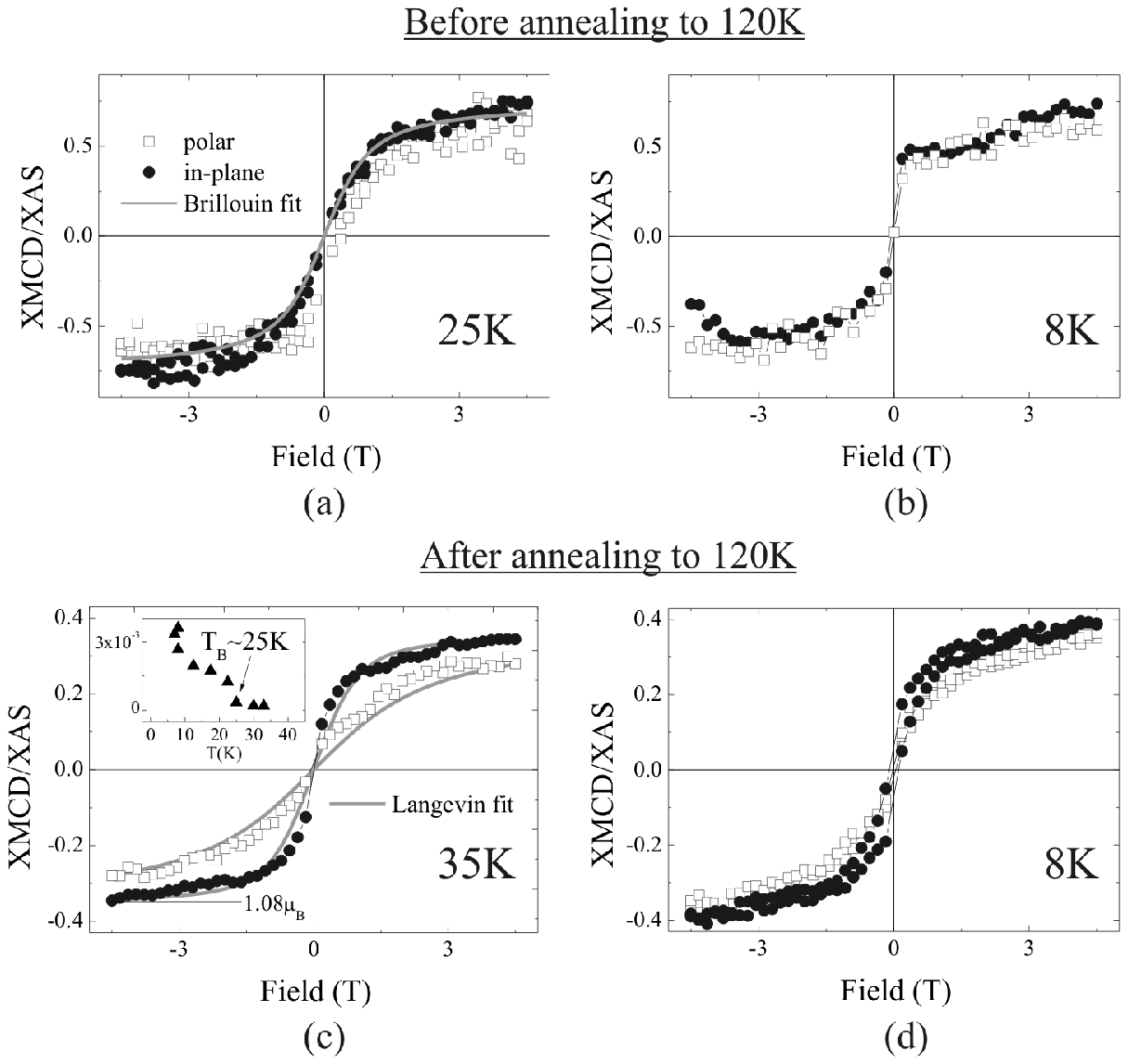}
  \caption{\label{CoAg2} Magnetic characterization of 0.05ML Co/50L
Xenon/Ag(111) before (a-b) and after (c-d) annealing at 120K; (a-b): hysteresis loops at 25K and 8K taken right after Co deposition
on the Xe buffer layer; (c-d): hysteresis loops at 35K and 8K taken after annealing the sample to 120K with consequent desorption
of the bulk Xe, cluster growth and contact with the substrate. The inset in (b) shows quasi-remanence measurements at $B$=5mT; the
full line in the hysteresis plot corresponds to a Langevin fit according to Eq. [5] in Ref.~[\onlinecite{Bornemannfit}].}
\end{figure*}

\label{Magnetism} Magnetic properties like spin and orbital moments, and in particular the magnetic anisotropy, are known to be
highly sensitive to clustersize, geometry, and interface effects. The XMCD technique provides surface sensitive information on
these magnetic parameters and therefore allows to extract information on the sample structure. We studied the magnetic properties
of the Cobalt clusters during classic BLAG with $12\pm1$ layers of Xenon and a Cobalt coverage of 0.05ML. Details about the
experimental procedure and the evaluation of the XMCD data are given in the appendix. \newline \indent Fig.~\ref{CoAg2}(top) shows
hysteresis loops taken shortly after Co deposition at two different temperatures.  The ordinate is the XMCD/XAS ratio evaluated at
the peak of the $L_{3}$ edge plotted versus magnetic field. To first approximation this ratio is proportional to the magnetization,
XMCD/XAS~$\propto M$, where $M$ is the average magnetization of the sample. We find that none of the magnetization loops in (a, b)
show remanent magnetization at zero field. In addition, the loops are isotropic, i.e. they do not show any difference between polar
($\phi = 0^{\circ}$) and in-plane ($\phi = 70^{\circ}$) magnetic field directions, where $\phi$ is the angle of the field with the
substrate normal. Reasons for this apparent absence of magnetic anisotropy could be an amorphous cluster structure, a random
distribution of the cluster easy axis or even a partial realignment of the clusters on the Xe layer in the presence of a torque
created by the magnetic field. A fit to the in-plane magnetization loops with a standard Brillouin function yields a spin block
size $N = 28\pm5$ atoms.\newline \indent The scenario changes upon desorption of the Xenon when the sample is annealed at 120K. In
Fig.~\ref{CoAg2}(c) and (d) we show hysteresis loops taken after Xe desorption and cooling the sample back to 35K and 8K,
respectively. While the Cobalt moments show no sign of anisotropy on the Xe layer, anisotropy emerges when the clusters make
contact with the surface. The in-plane direction now is an easy axis and we can estimate the value of the MAE via a
superparamagnetic fit using the procedure described in Ref.~[\onlinecite{Bornemannfit}] assuming an uniaxial symmetry concerning
the magnetic anisotropy. We obtain a hard axis in the polar direction with a MAE of $-0.15\pm 0.1$meV/atom and a spin block size of
$N=52\pm5$ atoms. The latter value is slightly larger than the cluster size of about 20-40 atoms found in Section \ref{Cluster
growth and morphology}. However, within the experimental error there is still good agreement, an indication that the thermodynamic
model employed to describe the nanoclusters magnetization is reasonable for our system. Further, the blocking temperature of the
clusters is obtained from temperature dependent measurements of the remanent magnetization at very small applied fields of $B =
5$mT shown in the inset of Fig.~\ref{CoAg2}(c). From the temperature at which the remanence disappears we find $T_{\text{B}}=
25\pm5$K.\newline As discussed in detail in the appendix in section~\ref{Appendix: Experimental} the XMCD technique allows for a
quantitative analysis of Co orbital and spin magnetic moments using the sum rules for 3$d$ metals. The moments per $d$-band hole
derived from the XMCD data are summarized in Table~\ref{tab:table1}. \newline We want to point out that in the case of Co situated
on thick Xe layers the signal-to-noise ratio in the XMCD was insufficient for a quantitative evaluation of the moments. Thus, in
this case only the XMCD/XAS ratio at the $L_3$ absorption edge is shown in Tab. \ref{tab:table1}.
 \begin{table}
  \caption{\label{tab:table1} Magnetic properties of 0.05ML Co/ 50L Xe/Ag(111)
  before and after Xe desorption (annealing to 120K). The average magnetic moments given in units of $\mu_{\text{B}}$ have
  been calculated from the saturated XMCD data at $T=8$K using the sum rules, with the magnetic field in-plane and polar
  with respect to the surface normal. The value of the
  XMCD/XAS ratio at the $L_3$ absorption edge is also indicated.}
  \begin{ruledtabular}
  \begin{tabular}{llllll}

    BLAG stage & $\mu_{\text{L}}\over{n_{\text{h}}}$ & $(\mu_{\text{S}}+7\mu_{\text{T}})\over{n_{\text{h}}}$ & $ \text{XMCD}\over{\text{XAS}}$\\
    \hline
     before Xe desorption \\
     \hline
 in-plane (8K) & - & - & 0.74$\pm0.05$\\
 polar (8K) & - & - & 0.64\\
 \hline
 after Xe desorption\\
 \hline
in-plane (8K) ¥ &0.08 ¥& 0.46 ¥ & 0.40¥ \\
polar (8K) &0.07¥ &0.33 ¥ & 0.36¥ \\

\end{tabular}
\end{ruledtabular}
\end{table}

\section{Buffer-layer assisted growth with atomically thin Xenon layers}
\label{Buffer-layer assisted growth with atomically thin Xe layers}

\subsection{Co cluster  morphology on Ag(111) and Pt(111)}
\label{Co cluster  morphology on Ag(111) and Pt(111)}

\begin{figure}[htbp]
  \centering
  \includegraphics[width=.45\textwidth]{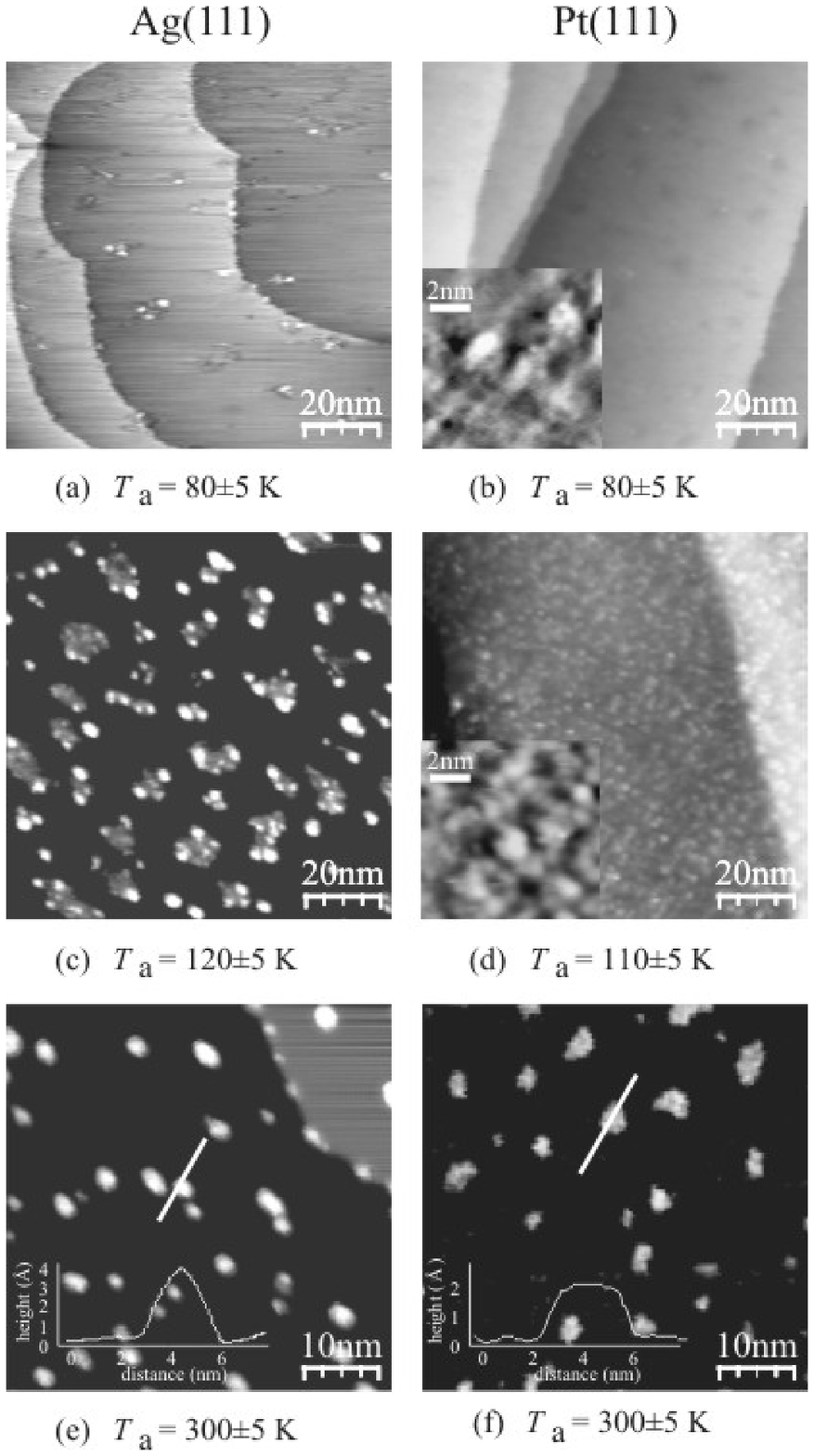}
\caption{\label{VT-STM2} STM topographies for Co/5L Xe/Ag(111) (left column) and Co/5L Xe/Pt(111)
  (right column) taken after annealing at $T_{a}$ indicated in the figure caption. The scanning temperature was 20K
   for (a)-(d) and 300K for (e) and (f). The inset in (e) and (f) shows the line scan across a Co island supported
   on the two different substrates.}
\end{figure}

In order to address the question if atoms or clusters interact with the substrate long before the last monolayer of the noble gas
is desorbed we studied the extreme case of BLAG with only a single layer (5 Langmuir) of Xenon. As substrates we used Ag(111) and
Pt(111). The Co coverage was again 5\% of an epitaxial monolayer, in analogy to the samples in the previous section.\newline Xe
grows on Pt(111) as a commensurate layer below coverages of $\Theta\simeq 0.33$ ML. At higher coverages a transition from
commensurate to incommensurate occurs that leads to a final hexagonal incommensurate rotated 2D solid phase~\cite{Kern1988}. As for
Xe/Ag(111), also for Xe/Pt(111) the desorption is a zero-order transition~\cite{Widdra1998}. The Xe desorption temperature for 1 ML
Xe/Pt(111) is about 100-107K, which is significantly higher than on Ag(111), due to the higher binding energy~\cite{Widdra1998}.
\newline \indent STM images taken on 0.05 ML Co / 1 ML Xe / Ag(111), as well as on
0.05 ML Co / 1 ML Xe / Pt(111), in the temperature range between 80K and 300K are summarized in Fig.~\ref{VT-STM2}. At low
temperatures, when the complete Xe monolayer is still adsorbed and cluster formation is in its early stage, we can hardly
distinguish the clusters from the Xe background (see Fig.~\ref{VT-STM2}(a) and (b)). \newline In the case of Ag(111)
(Fig.~\ref{VT-STM1}(a)) one observes streaks in the STM which points to a displacement of mobile Co clusters or Xe atoms by
field-induced diffusion while scanning the tip over the surface. Terrace step edges as well as defects and vacancies in the Xe
layer provide an energetically favorable position where the Co clusters are more strongly bound and therefore immobile.\newline For
the BLAG on Pt(111), we observe an increase of the apparent height corrugation $\Delta h$ when warming the sample from 80K ($\Delta
h \approx0.2-0.4${\AA}) to 110K ($\Delta h\approx0.6${\AA}) and finally to 150K ($\Delta h\approx1.6${\AA}). Our conclusion from
these and many other STM images is that the clusters are actually buried in the Xe buffer layer and become exposed as the Xe layer
desorbs.\newline On the Ag(111) surface we see a similar behavior as observed in the previous paragraph for thicker Xenon films at
comparable temperatures (Fig.~\ref{VT-STM1}(e)), that is residual Xe on the surface is pinned at Co clusters up to 150K.
\newline For the case of Co/Xe/Pt(111) due to the very small size of the Co clusters and their high density, it is difficult to
distinguish them from Xe-atoms and -clusters adsorbed on the surface. Nevertheless, we can observe an increase of the apparent
height corrugation when moving from 80K ($\Delta h\approx0.2-0.4${\AA}) to 110K ($\Delta h\approx0.6${\AA}) and finally to 150K
($\Delta h\approx1.6${\AA}). This is an indication that for the case of Pt(111) the clusters were buried in the Xe buffer layer and
become 'visible' after Xe removal. Further increase of the temperature up to RT (Fig.~\ref{VT-STM2} (e) and (f)) produces in both
cases an increase of the cluster size accompanied by a reduction of the density as observed in Fig.~\ref{VT-STM1}(f). \newline For
Ag(111) the average cluster size estimated from the STM data is now $N=16\pm 5$ at 140K and $N=44\pm 5$ at RT. On the Pt(111)
surface we have islands with an average number of atoms that changes from $N=6\pm 2$ at 120K to $N=62\pm 3$ at RT. We can conclude
that using constant BLAG parameters the system dependent growth dynamics can lead to structures with substantially different
properties on the two substrates: small compact 3D structures on the Ag(111) (Fig.~\ref{VT-STM2}(d)) and monolayer islands for the
Pt(111) (Fig.~\ref{VT-STM2}(e)).\newline \indent We conclude from these experiments that the substrate does have considerable
influence during buffer layer assisted growth and contributes to the final size, shape and distribution of the clusters, in
contrast to what has been reported so far.

\subsection{Comparison of the magnetism of Co clusters on Ag(111) and Pt(111) substrates}
\label{Magnetism2}

\begin{figure*}[htbp]
\centering
\includegraphics[width=.7\textwidth]{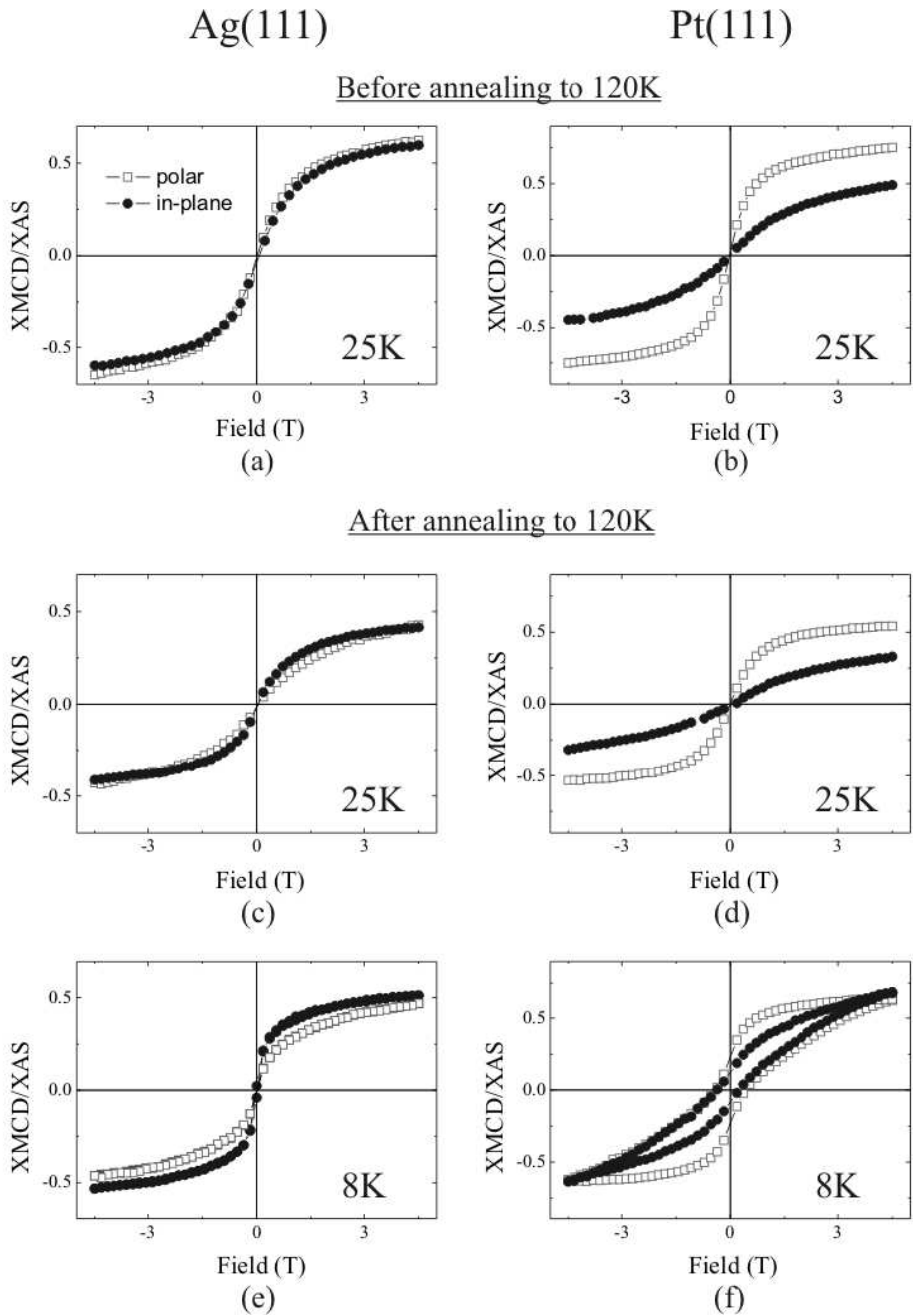}
\caption{\label{CoXeAgandCoXePt} Magnetic characterization of the systems in Fig.~\ref{VT-STM2}: 0.05ML Co/ 5LXenon/Ag(111) (left
column) and 0.05ML Co/ 5LXenon/Pt(111) (right column): (a) and (b) before and (c-f) after annealing at 120K, measured at 25K (c-d)
and 8K (e-f).}
\end{figure*}

The magnetism of all samples in this section~\ref{Buffer-layer assisted growth with atomically thin Xe layers} have been
investigated by XMCD. We have taken angular dependent X-ray absorption spectra at the Co L$_{3,2}$ edges, as a function of
temperature and magnetic field. The clusters considered in the following have been synthesized by BLAG using 0.05ML Co and 3-4ML
Xe.\newline \indent The complete set of magnetic moments deduced from the XMCD data are shown in Table~\ref{tab:table2} for the two
samples made on the Ag(111) and Pt(111) substrates, before and after annealing to 120K. The XMCD was measured at magnetic fields of
$B=4.5$T. The hysteresis loops obtained on those samples along the sample normal and under grazing incidence are summarized in
Fig.~\ref{CoXeAgandCoXePt}, again before and after Xe desorption. For clusters situated on Xenon/Ag(111) the magnetization is
isotropic (Fig.~\ref{CoXeAgandCoXePt}(a)), in analogy to what is observed in Fig.~\ref{CoAg2} for the case of thick buffer layers.
On the other hand clusters on Xenon/Pt(111) show a strikingly pronounced magnetic easy axis in the polar direction
(Fig.~\ref{CoXeAgandCoXePt}(b)), which endorses our STM interpretation of a stronger tendency of Cobalt to penetrate the Xenon.
After annealing both samples at $T=120$K and measuring again at 25K and 8K (Fig.~\ref{CoXeAgandCoXePt}(c) and (d)) in both cases
Ag(111) and Pt(111) a magnetic anisotropy is visible in the magnetization curves: while Co/Pt(111) still has a strong polar easy
axis, Co prefers the in-plane direction in the case of the Ag(111) substrate.\newline \indent The MAE was estimated in the
superparamagnetic regime from the Co magnetization loops on Pt(111) at 25K, and on Ag(111) at 8K, using the procedure described
earlier. We obtain a MAE of +0.67~meV/atom for Co/Pt, and -0.10~meV/atom for Co/Ag, respectively. An additional average induced
moment per Co atom due to the three Platinum nearest neighbors atoms $m_{\text{Pt}}=0.15\mu_{\text{B}}$/atom~\cite{Ebert2007} was
included in the calculation. For the spin block size we then get $N=18\pm5$ for Pt(111) and $N=15\pm4$ for Ag(111). Also in this
case the agreement with the cluster size found by STM measurements in Section \ref{Co cluster  morphology on Ag(111) and Pt(111)}
is quite acceptable. Remanence is observed at lowest temperatures of 8K in the case of the Pt(111) substrate due to the strong
polar MAE which pushes the blocking temperature to higher values.
\begin{table}
  \caption{\label{tab:table2} Magnetic properties for the samples Co/5L Xe/Ag(111) and Co/5L Xe/Pt(111) before and after annealing at 120K.
  The average magnetic moments given in units $\mu_{\text{B}}$ have been calculated from the saturated XMCD data at the indicated temperatures using the sum rules,
  with the magnetic field in-plane and polar with respect to the surface normal. The value of the
  XMCD/XAS ratio at the $L_3$ absorption edge is also indicated.}
\begin{ruledtabular}
\begin{tabular}{lllll}

   sample & $\mu_{\text{L}}\over{n_{\text{h}}}$ & $(\mu_{\text{S}}+7\mu_{\text{T}})\over{n_{\text{h}}}$ & $ \text{XMCD}\over{\text{XAS}}$\\
    \hline
    before Xe desorption\\
    \hline
Ag(111) in-plane (25K) ¥ & 0.10 ¥ & 0.53 ¥ & 0.60$\pm0.05$¥  \\
Ag(111) polar (25K)¥ & 0.11¥ & 0.60 ¥ & 0.65¥ \\
Pt(111) in-plane (25K)¥ & - ¥ & - ¥ & 0.45¥ \\
Pt(111) polar  (25K)¥ & 0.16 ¥ & 0.54 ¥ & 0.76¥  \\
\hline
 after Xe desorption\\
 \hline
Ag(111) in-plane (8K) ¥ & 0.11 ¥ & 0.49 ¥ & 0.51¥  \\
Ag(111) polar (8K)¥ & -¥ &  -¥ & 0.46¥ \\
Pt(111) in-plane (8K)¥ & - ¥ & -  ¥  & 0.38¥ \\
Pt(111) polar (8K)  ¥& 0.12¥ &  0.57¥ & 0.61¥  \\
\end{tabular}
\end{ruledtabular}
\end{table}

\section{$\textit{AB-INITIO}$ calculations of magnetic $\text{Co}_{N}$ islands on $\text{Ag(111)}$ and $\text{Pt(111)}$}

In order to interpret the experimentally observed trends in the magnetic properties of these deposited Co clusters we performed
{\it ab-initio} calculations within the local density approximation of density functional theory (DFT)~\cite{VWN80}, using the
spin-polarized relativistic Korringa-Kohn-Rostoker multiple scattering formalism~\cite{Ebe00}. In this scheme, the Dirac Green's
function was calculated self-consistently for large single and multi-layer Co islands assuming pseudomorphic deposition on a
38-layer Ag (Pt) slab having the experimental lattice constant of 4.085 (3.924)~\AA. Furthermore we applied the atomic sphere
approximation to the potentials and neglected lattice relaxations (see Ref.~[\onlinecite{BMS+07}] for more details). The influence
of coordination effects on the magnetism of the deposited Co clusters were studied by a systematic increase of the island size.
\newline \indent Table~\ref{tab:table3} shows calculated spin ($\mu_{\text{S}}$), orbital ($\mu_{\text{L}}$) and intra-atomic dipolar moment
($\mu_{\text{T}}$) and magneto-crystalline anisotropy energy (MAE). For the system Co/Pt(111) we find an out-of-plane easy axis
direction at every island size with the exception of Co$_3$ and Co$_7$ where the anisotropy is slightly in-plane but with a value
close to zero. These exceptions reflect oscillations of the magnetic anisotropy for the smallest cluster sizes. More interestingly
for Co/Ag(111) we also find a strong (10.98meV/atom) out-of-plane anisotropy for the single atom case, which is highly sensitive to
a lateral coordination with other Cobalt atoms: a sudden easy axis reorientation to the in-plane direction is predicted for the Co
dimer case accompanied by an abrupt drop in the MAE absolute value (-1.23meV/atom). After that, adding more atoms leaves the easy
axis direction in-plane and the MAE varies only slightly with increasing number of atoms in the first layer, until reaching the ML
value of -1.62meV/atom. Stacking of Co atoms in a second layer reduces the absolute MAE value dramatically (almost a factor of 10),
but does not change the easy axis direction for islands with more then 1 atom. The positive MAE in the case of Co$_4$ (three atoms
in the first layer and one atom in the second) is due to the single Co atom in the second layer. A third layer of Co does not
substantially change the situation anymore. Concerning the spin and orbital moments, we observe that they both decrease if the
number of atoms in the island is increased, an effect known also from the literature \cite{Sipr,Ebert2007}. The intra-atomic
dipolar term decreases progressively in absolute value with increasing number of layers, indicating that the distribution of the
magnetic moment is becoming more isotropic when the cluster is becoming more compact.

\begin{table}
  \caption{\label{tab:table3} {\it Ab-initio} calculations for the systems Co/Pt(111) and Co/Ag(111) calculated for ML height Co island
  with increasing number of atoms and for bi-layer and tri-layer height islands. In the table average values for magnetic anisotropy
  (in meV per atom) and magnetic spin, orbital and intra-atomic dipolar moment (in $\mu_{\text{B}}$ per atom) are reported.
   Positive (negative) values for the MAE indicate out-of-plane (in-plane) easy axes.}

  \begin{ruledtabular}
  \begin{tabular}{lllll}
sample & MAE & $\mu_{\text{S}}$ & $\mu_{\text{L}}$ & $\mu_{\text{T}}$ \\
\hline\\
ML cluster/Pt(111)\\
\hline
Co1&4.88&2.269&0.604&-0.209\\
Co2¥&2.24&2.160&0.441&-0.045\\
Co3¥&-0.12&2.081&0.234&-0.088\\
Co7¥&-0.25&2.024&0.192&-0.025\\
Co19&0.22&1.967&0.168& \\
Co37¥&0.24&1.947&0.160&-0.047\\
 \hline
 \\

ML cluster/Ag(111)\\
    \hline
Co1&10.98&2.145&1.350&-0.023\\
Co2&-1.23&2.050&0.506&¥ 0.012\\
Co3&-2.71&2.009&0.410&¥ 0.012\\
Co7&-2.91&1.954&0.243&¥ 0.030\\
Co19&-1.60&1.915&0.221&-0.015\\
Co37&-1.78&1.899&0.214&-0.018\\
monolayer&-1.62&1.872&0.186&-0.027\\
\hline\\

bi-layer cluster/Ag(111)\\
    \hline
Co4&¥ 0.70&2.022&0.571&¥ 0.005\\
Co10&-0.235&1.921&0.277&¥ 0.016\\
Co31&-0.34&1.895&0.217&-0.006\\
Co64&-0.234&1.884&0.209&-0.007\\
\hline\\

tri-layer cluster/Ag(111)\\
    \hline
Co39&-0.42&1.899&0.221& 0.003\\
Co82&-0.21&1.888&0.210&0.0002\\

  \end{tabular}
\end{ruledtabular}
\end{table}

\section{Discussion}
\label{Discussion)>>}

\subsection{Growth dynamics of Co nano-structures on Xenon buffer layers}
\begin{figure}
\includegraphics[width=.45\textwidth]{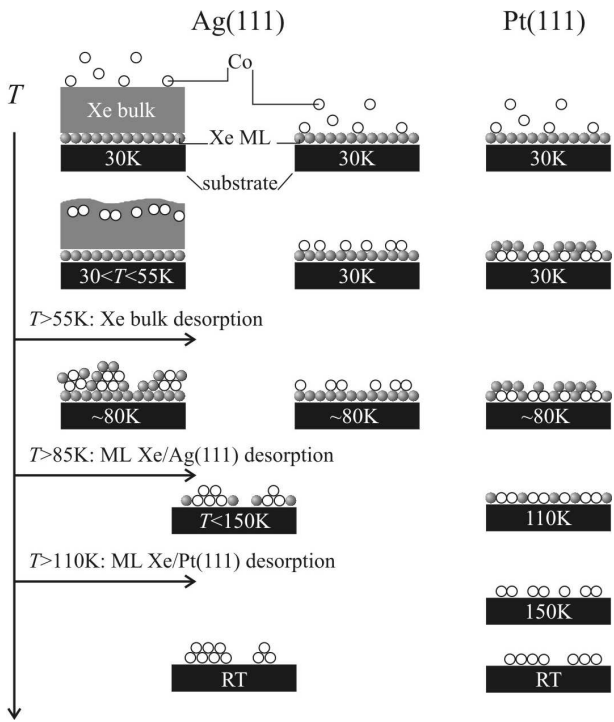}
\caption{\label{model} Detailed scheme of the cluster
formation dynamics for the three different samples: Co/50L
Xe/Ag(111)(left), Co/5L Xe/Ag(111)(center),
and Co/5L Xe/Pt(111)(right). The arrow on the left indicates the
raising temperature $T$ from top to bottom.}
\end{figure}
Key observations during our study of the growth of Co on thick and thin Xe layers from the presented VT-STM experiments are (i) the
observation of delayed Xe desorption in the vicinity of the Co clusters, (ii) differences in the cluster-substrate interaction for
Ag and Pt substrates, resulting in a considerably stronger tendency for clusters to penetrate the Xe matrix for the case of Pt
substrates, (iii) differences in the cluster morphology on Ag and Pt substrates, as well as cluster ripening during annealing of
the samples to room temperature. In this paragraph we will now discuss the details of Xe desorption and the cluster formation in
more detail.\newline \indent One of the most widely used concepts for thin film and cluster growth is the consideration of surface
or interface energies of substrates and adlayers~\cite{Miedema1981}. The interfacial energy $\gamma^{0}_{AB}$ at the interface of
two materials $A$ and $B$ is commonly expressed as $\gamma^{0}_{AB}= \gamma^{0}_{A}+\gamma^{0}_{B}+\Delta\gamma^{\text{ads}}_{AB}$,
where $\gamma^{0}_{i}$ is the surface free energy of material $i$ and $\Delta\gamma^{\text{ads}}_{AB}$ the interfacial adhesion of
the system $AB$. The $\Delta\gamma^{\text{ads}}_{AB}$ can be calculated (at zero temperature) for metal/Xe and metal/metal systems,
using values for the surface free energy and interfacial adhesion from Ref.~[\onlinecite{Miedema1981}]. We obtain interfacial
energies $\gamma^{0}_{AB}$ of 2.33J/m$^2$, 0.23J/m$^2$, and 0.007J/m$^2$ for Co/Xe, Co/Ag and Co/Pt, respectively. The gain in free
energy upon formation of an interface between two materials can be written as: $\Delta\gamma = \gamma^{0}_{A} + \gamma^{0}_{AB}-
\gamma^{0}_{B}$. This formula simply expresses a comparison between the total free energy of the system in two different states: an
initial state where the elements  $A$ and $B$ are both exposed to vacuum and a final state where $A$ is wetting $B$, and therefore
the $AB$ interface substitutes the $B$-vacuum interface. Negative or very small values of $\Delta\gamma$ indicate that the
formation of the interface is favored energetically and therefore a layer by layer growth is expected. In our case, for $A$=Co we
found the values 1.53J/m$^2$ and 0.007J/m$^2$ for the two substrates $B$=Ag and Pt, respectively, that is Co wets Pt(111) but not
Ag(111). Concerning the Xe/Co interface, one finds $\Delta\gamma=-0.188$J/m$^2$ for Xe wetting the Co surface ($A$=Xe and $B$=Co).
As already discussed in~\cite{Weaver2008} for the case of Au nanoparticles on Xe, these results suggest that the Xe will coat the
Co nanoparticles in order to minimize the surface energy.\newline Similar energy considerations can also be used to predict the
morphology of the clusters in equilibrium. The Young-Dupr\'{e} formula for a liquid droplet on a solid relates the contact angle
$\theta$ of a droplet to the surface free energies of the solid and liquid, and the interfacial energy of the solid-liquid system:
$\gamma^{0}_{\text S}=\gamma^{0}_{\text{SL}}+\gamma^{0}_{\text L}\cos\theta$. Special cases are complete wetting for
$\theta=0^{\circ}$, and a crossover from wetting to dewetting for $\theta=90^{\circ}$.\newline Evaluating the Co cluster morphology
with this approach predicts total wetting for Co/Pt ($\theta=0^{\circ}$), partial wetting for Co/Ag ($\theta=66.4^{\circ}$), and
dewetting for Co/Xe ($\theta=154^{\circ}$). According to this estimate, the formation of a Co/metal substrate interface seems
energetically most favorable, which could potentially destabilize 3-layer Co/Xe/metal system. Similar arguments have been used
recently to explain experimentally observed coating of Au nanoclusters by Xe leading to embedding in a Xe buffer
layer~\cite{Weaver2008}. Thus, the analysis of the surface free energies explains very well our experimental observation of the
formation of hemispherical Co clusters on Xe/Ag, and the differences of the cluster morphology after making contact with the Ag and
Pt surface. Especially on the Pt(111) the strong tendency towards complete wetting results in the formation of monoatomically flat,
epitaxial islands.\newline \indent For a better understanding of the temperature dependent dynamics during BLAG it is instructive
to compare the desorption energies of Xe in different environments. The desorption energy for bulk Xe is 161meV per Xenon
atom~\cite{Leming1970}, while for a single Xe monolayer on on Ag(111) the desorption energy is 208meV/atom~\cite{Igarashi2003}, and
286meV/atom for 1ML of Xe on Pt(111)~\cite{Widdra1998}. The higher desorption energy for single Xe monolayers is due to
interactions with the supporting substrate. It is therefore reasonable to assume that the desorption of the Xe is a two-step
process. At the bulk desorption temperature, the entire Xe layer will evaporate except of the first monolayer that is in direct
contact with the metal substrate. Higher temperatures are required to break the bonds between the Xe monolayer and the substrates,
and the Xe desorption is delayed to even higher temperatures in the vicinity of surface defects or clusters at the surfaces. A
similar pinning effect has also been reported for Ag clusters, soft-landed on a Kr buffer layer~\cite{Harbich2001}. As a matter of
fact, the required evaporation heat per atom for a single Xe monolayer is determined by intralayer Xe-Xe- interactions, as well as
Xe-substrate interactions. There are two different contributions to the Xe-Xe intralayer interactions, which are attractive
van-der-Waals interaction and a repulsive contribution originating in the interaction between Xe and the substrate. The total
lateral binding energy for 1ML Xe/Ag(111) was reported to be 54.37meV/atom~\cite{Igarashi2003}. Strain effects or the formation of
induced dipoles in the Xe layer, for instance, can in fact increase the described repulsive contribution, hence weakening the total
Xe-Xe interaction. The measured surface dipoles for Xe/Ag(111) is 0.2D~\cite{Roberts1976}, which corresponds to an average
repulsive contribution of about 7meV/Xe atom. For Xe/Pt(111), the induced dipole moment in the Xe layer is considerably higher,
namely 0.53D, corresponding to an increased repulsive energy per Xe atom of 19meV~\cite{Schönhense1986}. The higher surface dipole,
together with other effects such as a higher work function, an unfilled $d$ band, and stronger corrugation of the Pt(111) surface
potential, contributes to a total lateral binding energy per Xe atom which is reduced by 30meV/atom, compared to
Ag(111)~\cite{Kern1988, Gottlieb1991}.\newline \indent We argue here that the observation of Co cluster embedding in the Xe layer
is a result of the described weakened Xe-Xe bonds, and the presence of attractive van-der-Waals interaction between the Co and the
metal substrate across  the Xe buffer layer. For otherwise identical geometry the strength of the van-der-Waals interaction also
depends on the dielectric function of the materials~\cite{Israelachvili1972}. The presented comparison of BLAG on Pt and Ag
substrates thus shows that the substrate can have a significant influence on the final size and shape of the clusters, as it
determines the bond strength in a thin buffer layer, cluster-substrate interactions as well as the wetting behavior. As a result of
the differences in BLAG, Co clusters on Pt(111) tend to be smaller and of flat shape, in comparison to Ag(111) substrates where the
same BLAG parameters result in larger clusters of hemispherical shape. \newline The above discussion is summarized in the growth
model displayed in Fig. \ref{model}.

\subsection{Analysis of the magnetic properties during Co
self-assembly - Comparison to {\it ab-initio} theory}

In this section the measured magnetization as well as evaluated orbital and effective spin moments of the Co nanostructures during
the different steps of BLAG (Table~\ref{tab:table1}-\ref{tab:table2}) are discussed in more detail since they allow for a
correlation with the morphology information extracted from STM. Table~\ref{tab:table4} shows an overview of magnetic moments in
relevant geometries calculated with {\it ab-initio} theory together with the values obtained from the XMCD sum rules.\newline
\indent First qualitative trends can be seen in the XMCD/XAS values at magnetic fields $B=4.5$T given in
Table~\ref{tab:table1}-~\ref{tab:table2}. It is a well-known effect that spin and orbital moments are reduced as 3$d$ clusters grow
in size. Indeed, the XMCD/XAS values reflecting the average magnetization $M$ is always smaller after Xe desorption, when the
cluster size $N$ is increased. When comparing $M$ before Xe desorption then among the three measured samples Cobalt on 12ML of Xe
possesses the highest XMCD/XAS value of $\approx 0.7$ in average but shows no magnetic anisotropy. This is coherent with the
picture that smallest clusters on a thick Xe layer are electronically decoupled from the substrate. The same amount of Co on only
3.5ML of Xe on Ag(111) shows already a slightly lower XMCD/XAS value of $\approx 0.6$ in average but still no MAE. Finally the
sample with 3.5ML Xe on Pt(111) shows a similar value of XMCD/XAS$\approx 0.6$ in average but a rather strong MAE.
\begin{table*}

  \caption{\label{tab:table4} Average magnetic moments per atom in units $\mu_{\text{B}}$ for 0.05ML Co on different
  substrates measured in the direction of the easy axis in a magnetic field of $B=4.5$T. Expected values from theory for
 epitaxial monolayer and bi-layer islands are shown in the lower part of the table. The number of holes in the $d$ band was
  assumed to be $n_{\text{h}}=$2.49. Magnetic anisotropy energies are given in units of meV.}

  \begin{ruledtabular}
  \begin{tabular}{llllll}

    sample & $\mu_{\text{L}}/\text{atom}$ & $(\mu_{\text{S}}+7\mu_{\text{T}})/\text{atom}$ & \text{anisotropy/atom} & \text{spin block size} $N$ & \text{STM size}\\
    \hline
    \text{experiment} \\
        \hline
        Ag(111)-50L Xe\\
before Xe desorption (25K)  ¥ &   ¥ &  ¥& isotropic& 28\footnote{after fitting the magnetization curve with a standard Brillouin function}$\pm 5$& \\
after Xe desorption (8K) &  0.20$\pm 0.1$ & 1.14$\pm 0.1$ ¥ & &\\
after Xe desorption (35K) &  0.21 & 0.87 ¥ &-0.15$\pm0.1$ & 52 $\pm 5$& 40$\pm 7$\\
\hline
Ag(111)-5L Xe\\
before Xe desorption (25K)  ¥ &  0.28 ¥ & 1.49 ¥& isotropic& 22$\pm 4$& \\
after Xe desorption (8K) & 0.27¥ &  1.22¥ &-0.10 & &\\
after Xe desorption (25K)& 0.27& 0.93& & 15$\pm 4$&16$\pm 5$ \\
 \hline
Pt(111)-5L Xe\\
before Xe desorption (25K) ¥ & 0.39 ¥ & 1.35 & 0.59 & 17$\pm5$& \\
after Xe desorption (8K) ¥& 0.29¥ &  1.43  & - & - & \\
after Xe desorption (25K) ¥& 0.25¥ &  1.20  & 0.67 & 18$\pm5$&6$\pm 2$ \\
    \hline
theory  & $\mu_{\text{L}}/\text{atom}$ & $\mu_{\text{S}}/\text{atom}$ & MAE/atom \\
    \hline
Co10/Ag(111) bi-layer ¥ & 0.277¥ &  1.921¥ & -0.235 & 10\\
Co31/Ag(111) bi-layer ¥ & 0.217¥ &  1.895¥ & -0.34 & 31\\
Co19/Pt(111) monolayer&0.168&1.967& 0.22 & 19\\
Co37/Pt(111) monolayer ¥&0.160&1.947& 0.24 & 37\\
\hline Co bulk \cite{Chen}¥&0.154&1.62&  &
 \end{tabular}
\end{ruledtabular}
\end{table*}
\newline In the following we want to address separately the
results of orbital moments which are related to magnetic anisotropy effects, and trends in the spin moments.

{\it Magnetic anisotropy and orbital moments:} For a better understanding we briefly recall that the MAE is directly related to
anisotropies of the orbital moments via the spin-orbit coupling constant $\zeta$. In a simple case of a surface supported
semispherical Co cluster a uniaxial MAE $\propto \zeta (\mu_{\text L}^{\parallel}-\mu_{\text L}^{\perp})$ is expected, assuming
majority Co $d$-bands are filled~\cite{Bruno1989}. Here, $(\mu_{\text L}^{\parallel}-\mu_{\text L}^{\perp})$ is the anisotropy of
the orbital moments parallel and perpendicular to the surface induced by the asymmetric environment of Co. In general a reduction
of the Co orbital moments are related to the hybridization between Co $d$ states with $sp$ and $d$ bands of the substrate and to
internal $d$-$d$ hybridization within the cluster. The latter of course increases with the average cluster size. \newline \indent
In Table~\ref{tab:table3} the quenching effect on $\mu_{\text L}$ with increasing cluster sizes is also shown by DFT based
calculations. Hybridization effects from the substrate are expected to be smaller for Ag(111) compared to Pt(111), since in the
former case the full $d$ band is shifted far below the Fermi level. Again this is reflected in the {\it ab-initio} calculations,
where for a given cluster geometry the orbital moments are higher for the case of Ag(111). In the case of Co on Ag(111) the drop of
orbital moment is therefore mainly due to the internal $d$-$d$ hybridization of the Co with increasing number of Co-Co neighbors in
the cluster: the nanomagnets become larger upon desorption of the bulk Xe at 120K, which is seen both in STM and in the spin block
size $N$ (see Table~\ref{tab:table5}). Sample I prepared on Ag(111) using thick Xe layers shows a smaller orbital moment after
annealing to 120K compared to sample II prepared using a thin Xe layer. This suggests an increased cluster size with increasing
buffer layer thickness in agreement with both our STM investigation and the findings by Weaver {\it et al.}. We want to stress,
however, that the error bar in the orbital moment evaluation is relatively large. Nevertheless, also the absolute values
$\mu_{\text L}=0.27\mu_{\text B}$ per atom measured on Ag(111) are in good agreement with the calculations of bi-layer islands of
the experimentally derived cluster sizes $N=16$ (estimation from STM) and $N=15$ (spin block size): from Table~\ref{tab:table3} we
expect $\mu_{\text L}$ to be $0.22 - 0.28\mu_{\text B}$. In the case of Pt(111) {\it ab-initio} theory of monolayer islands with
spin block sizes $N=25-30$ underestimates $\mu_{\text L}$ by about $30\%$. However, using the cluster size $N\approx6$ estimated
from STM leads to a much better agreement between experiment and theory. In the case of the Co/Xe/Pt(111) system we can compare our
results to what was found by Gambardella {\it et al.}~\cite{GambardellaScience} for Co on Pt(111). A monolayer island of 7-8 atoms
has a MAE of about 1meV/atom and an orbital moment of about $0.35 \mu_{\text B}$/atom. Within the experimental errors, these
numbers agree well with that of BLAG-grown Co clusters. \newline \indent The well pronounced polar MAE observed already right after
Co deposition on the Xe layer as compared to the lack of anisotropy for Co/Xe/Ag(111) indicates that in the first case the Co atoms
and nanostructures cannot be considered as ´free´. In our view the effect can only occur in presence of a broken symmetry as
discussed in Ref.~[\onlinecite{GambardellaScience}], that means a chemical bond with the Pt(111) surface. In line with what is
discussed in the previous subsection on the STM results, we propose that Co penetrates the few Xe layers already at 25K to make
contact with the substrate.

{\it Spin moments:} The theoretical values of $\mu_{\text S}$ in Table~\ref{tab:table3} show that depending on the substrate the
spin moment is expected to decrease monotonously by up to 14$\%$ when going from Co$_{1}$ to Co$_{31}$ monolayer islands. In
agreement with this trend, the experimentally determined effective spin moments $(\mu_{\text{S}}+7\mu_{\text{T}})$ in
Table~\ref{tab:table4} are larger for the samples made with a thin buffer layer, where clusters are expected to be smaller.
However, absolute experimental spin moments $\mu_{\text S}$ (the contribution of the intra-atomic magnetic dipole moment
$\mu_{\text T}$ was accounted for using theory values in Table~\ref{tab:table3}) are smaller then those calculated for small
clusters. For example in the case of Co islands on Pt(111) we find spin moments of only $\mu_{\text S} = 1.76\mu_{B}$/atom compared
to the values $\mu_{\text S} = 1.96\mu_{B}$/atom predicted by the calculations. The experimental value is thus more comparable to
bulk values of $1.62 \mu_{B}$/atom~\cite{Chen}, where the spin moment is known to be reduced due to the large degree of Co-Co
coordination. For the samples made on Ag(111) we even find Co spin moments which are below the bulk value. When comparing the Co /
5L Xe / Ag(111) sample before and after Xe desorption we conclude that in the case of Ag(111) the particularly pronounced reduction
of the spin moment is mainly happening during Xe desorption. On the one hand this is expected due to the increase of cluster size,
but on the other hand we believe that part of the effect is due to the contact between Co and Ag(111). Otherwise it can not be
explained why the spin moments remain so much higher for the case of Co on Pt(111). The calculated spin values for ML cluster
geometries in Table~\ref{tab:table3} support the more pronounced quenching effect for Ag(111) substrates although the magnitude is
underestimated. In the following we want to discuss possible reasons for substrate induced quenching of average spin moments.
\newline \indent A quenching of the spin moment of Co in contact with a
non-magnetic metal has been found for Co nanoclusters embedded in a Cu matrix~\cite{Eastham,Sellmyer}. The authors could attribute
it to the cluster-matrix hybridization and presence of Rudermann-Kittel-Kasuya-Yosida (RKKY) type cluster-cluster interactions. We,
however, exclude this explanation in our case for two reasons. First, the average distances between clusters are larger then $2$nm
which makes the RKKY inter-cluster interaction negligible. To give an order of magnitude, for Co$_{32}$ clusters embedded in a Cu
matrix~\cite{Altbir} and cluster-cluster distances between $2$nm and $3$nm, calculated RKKY oscillations give interaction energies
between $0.03$meV and $0.005$meV, respectively. At experimental temperatures of $8$K ($0.7$meV) used in our experiments these
interactions should not play a role. The second argument is the trend of magnetization versus cluster size and density: the
magnetization is smaller for the sample made with 50L Xe, which corresponds to larger clusters with smaller cluster density.
Instead in Ref.~\cite{Sellmyer} it was found that, as a consequence of cluster-cluster interactions, the magnetization increases
with the cluster size and, for a given size, decreases with the cluster concentration.\newline \indent Screening effects due to
polarization of the Ag atoms surrounding the Co cluster is another possible explanation of the reduced spin moments. Recently, in
an experiment on Co nanoparticles embedded in a Ag matrix~\cite{Luis}, Ag atoms were shown to exhibit a non-vanishing dichroic
signal in an external magnetic field of 1Tesla. Although the induced Ag moments in the presence of Co atoms point in the same
direction as those of Co, charge transfer processes between Co and Ag need to be taken into account, which leads to incomplete
filling of the Ag \textit{d} bands and can decrease the average Co moments.
\newline \indent Finally we want to consider the possibility of a non-collinear
alignment of the Co atoms inside the cluster due to the interaction
with the substrate, which can lead to a reduction of the average
spin moments. Among other reasons such as geometrical frustration,
one of the reasons for non-collinear magnetism is the
Dzyaloshinski-Moriya (DM) term, also called 'anisotropic exchange
interaction'. This term is usually only important in the case of
weak exchange interactions between magnetic atoms, which is why it
is often not taken into account in most ferromagnetic systems.
Contrary to that, recently it was calculated that in presence of a
substrate with strong spin-orbit coupling~\cite{Mankovsky} a
non-collinear spin-structure could be stabilized even in presence of
strong ferromagnetic exchange interaction among atoms as in the case
of a Co dimer. In particular the authors suggest that DM couplings
can affect the spin structure around the edges of larger
nanostructures like those studied in this work.

\section{Conclusions}
\label{Conclusions)>>}

We have presented STM images and magnetic measurements on Co nanoclusters during different stages of Xe buffer layer assisted
growth. We find that Co clusters stay at the surface of Xe on Ag(111) substrates until making direct contact with the Ag(111). In
contrast on Pt(111) substrates, Co clusters have a strong tendency to penetrate Xe layers of a several monolayer thickness until
they get in contact with the substrate. We explain this behavior in terms of a complex interplay of Xe-Xe and Xe-substrate
interactions. Electronic decoupling of the clusters and the substrate is therefore only achieved on Ag(111) substrates, as
concluded also from the absence of magnetic anisotropy in this case. The magnetic anisotropy of clusters in contact with Ag and Pt
substrates is determined by the magnetic interface anisotropy and remanent out-of-plane magnetization is found for Co/Pt(111) at
temperatures below 25K. Cluster size effects and the contact with the substrate are also reflected in the spin- and orbital
magnetic moments. Trends obtained by XMCD could be reproduced in {\it ab-initio} DFT calculations. From the calculations it is
evident that magnetic properties like the orbital moments but especially the appearance of magnetic anisotropy is largely
determined by cluster-substrate interface effects.

\section{Appendix: Experimental}
\label{Appendix: Experimental}

\begin{table}
  \caption{\label{tab:table5} Overview of the experimental parameters used for the samples described in the text.}
  \begin{ruledtabular}
  \begin{tabular}{lllll}
        & STM  & & XMCD  \\
    sample label & Co & Xe  & Co & Xe   \\
    \hline
Ag(111)  ¥ & 0.05ML ¥ & 50L ¥ & 0.05ML  ¥ & 12ML \\
Ag(111)  ¥ & 0.05ML¥ & 5L ¥ & 0.08ML  ¥ &3-4ML \\
Pt(111)  ¥& 0.05ML ¥ & 5L ¥ & 0.06ML ¥ &3-4ML \\

  \end{tabular}
\end{ruledtabular}
\end{table}

\begin{figure*}[htbp]
  \centering
\includegraphics[width=.7\textwidth]{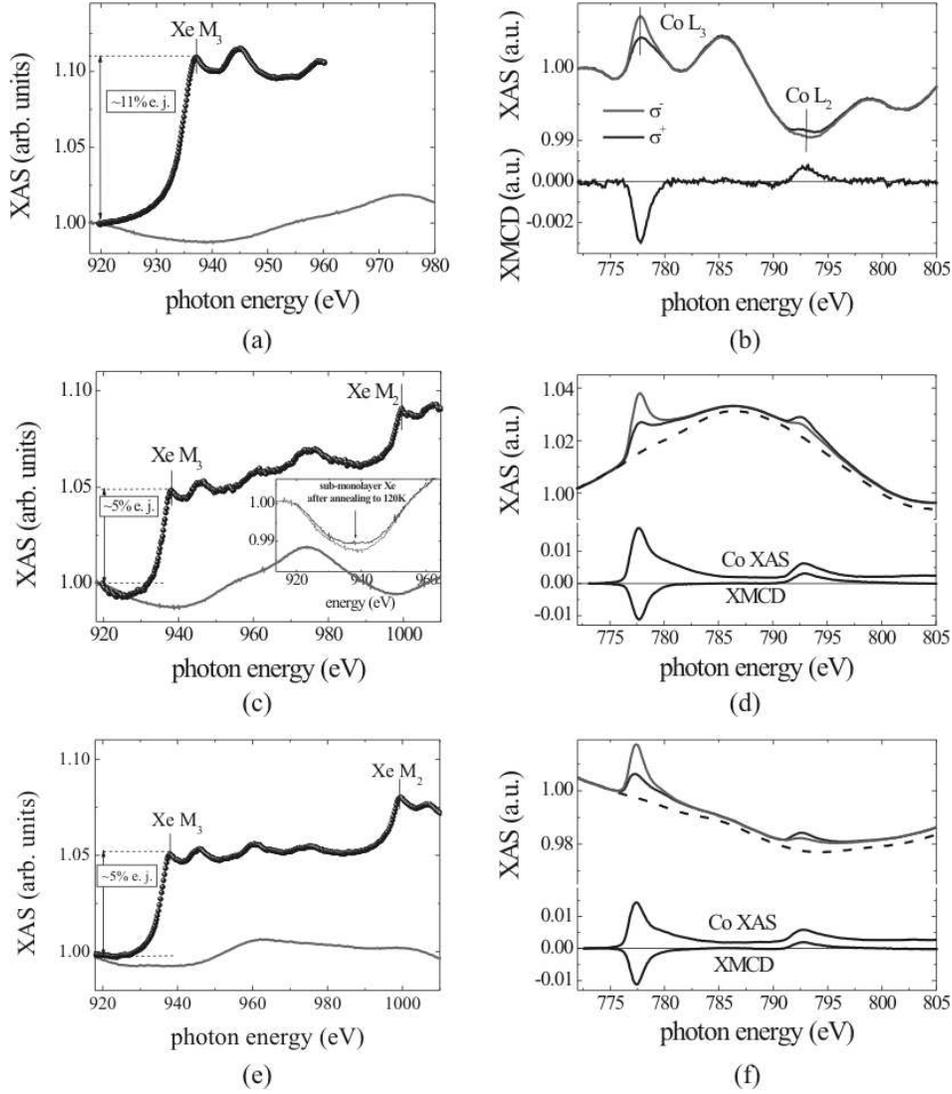}
\caption{\label{XenonXAS} XAS signal for the three samples described in the text,
  showing the sample composition and dichroism right after deposition of Xe and Co
at 25K: (a), (c), and (e): Xe $M_{3,2}$ edges for 12ML Xe/Ag(111) (a), 3-4ML Xe/Ag(111) (c) and 3-4ML Xe/Pt(111) (d); the
substrates background (green line) is also plotted; (b), (d), and (g): Co $L_{3,2}$ edges for 0.05ML Co/12L Xe/Ag(111) (b), 0.08ML
Co/3-4ML Xe/Ag(111) and 0.06ML Co/3-4ML Xe/Pt(111); dashed lines indicates the Xe/substrates background. The inset in (c) shows the
XAS taken after annealing at 120K for Xe/Ag(111) (blue line): a small amount of Xe (sub-monolayer coverage) is still present at
this temperature on the Ag(111) surface.}
\end{figure*}

VT-STM experiments were performed at the Max-Planck Institute for Solid State Research in Stuttgart, and XMCD experiments for
magnetic characterization were done at the European Synchrotron Radiation Facility (ESRF) in Grenoble, beamline ID08. In both cases
the samples were prepared and measured in UHV chambers with base pressure about $3\cdot10^{-10}$~mbar. A precise determination of
temperature, Xenon and Cobalt coverage as shown below ensured the comparability of the experiments.  \newline \indent The Cobalt
coverage was calibrated by a microbalance in the STM measurements and by the Co edge-jump at the $L_{3}$ edge for the samples used
in XMCD measurements. The Xenon coverage was calibrated from the partial pressure in the UHV chamber (converted to Langmuir and
then to film thicknesses in units of {\AA}ngstr\"{o}ms), for the STM measurements and based on the saturation of the TEY current
upon Xenon adsorption for the XMCD measurements (it will be described in more details in Ref.~\cite{SessiTh}). In
Table~\ref{tab:table5} we find a scheme with the experimental parameters used to fabricate the samples described in the paper. The
Ag(111) and Pt(111) substrates were cleaned by several sputtering/annealing cycles. Cleanliness and ordering of the crystal surface
was verified by Auger, LEED and STM measurements. No contamination from Oxygen and Carbon-monoxide were detected. During the XMCD
measurements the Oxygen contamination checks prior and after the measurements were done using x-ray absorption spectra (XAS) at the
Oxygen $K$-edge absorption line situated at 543.1eV. \newline \indent For all the experiments the Xenon was pre-adsorbed on the
sample at a temperature of about 30K. At the same temperature Co was deposited from an e-beam evaporator. For the morphologic
characterization, the sample was prepared in the manipulator and then transferred in the VT-STM, that was pre-cooled by liquid
Helium flow. For magnetic characterization, the clean crystals were transferred to the magnet chamber under UHV conditions with a
pressure of $1\cdot10^{-10}$mbar. In this case, adsorption of Xe and Co deposition were performed inside the high-field magnet
chamber at ID08.\newline \indent Before Co deposition, the XAS background was recorded at the energy range of the Co $L_{3,2}$ edge
for the systems of interest: Ag(111), Pt(111) as well as Xenon/Ag(111) and Xenon/Pt(111) for the respective samples under study. In
Fig.~\ref{XenonXAS}(a),(c), and (e) XAS spectra at the Xenon $M_{3,2}$ edge jump for 12 and 3-4 layers of Xe adsorbed on Ag(111)
and Pt(111) are shown. In Fig.~\ref{XenonXAS}(b), (d), and (f) we have the total electron yield signal for the three samples
recorded right after Co deposition on the Xe/substrate at $T = 25$K. The measurements have been done at $\sigma^+$ and $\sigma^-$,
in the Cobalt $L_{3,2}$ energy range and in a magnetic field of 4.5 Tesla. The background spectrum before Co evaporation is shown
as a dashed line, which was used to separate the Co XAS shown in the same graph below. Also shown is the XMCD signal of Cobalt
indicating a sizable magnetic moment.\newline \indent In the last step of BLAG the sample was annealed up to 120K in order to
desorb the Xe from the substrate, being the nominal desorption temperature $T_{\text{ des}}^{\text{substrate}}$ of the Xe monolayer
from $T_{\text{des}}^{\text{Ag}}\approx 80$K and $T_{\text{des}}^{\text{Pt}}\approx 110$K. The measurements were repeated also
after annealing the sample to 120K. As we can see from the inset in Fig.~\ref{XenonXAS}, a sub-monolayer amount of Xe was still
present on the substrate at these temperatures, due to the 'pinning' of Xe atoms on the Co clusters. After annealing at $120$K it
was verified that no Oxygen contaminations appeared. The XAS intensity was measured recording the total photoelectron yield as a
function of the x-ray energy for positive ($\sigma^+$) and negative ($\sigma^-$) x-ray circular polarization ($99 \pm 1 \%$ degree
of polarization). Magnetic fields up to $B=4.5$~T were applied parallel and antiparallel to the photon beam.  The angle of
incidence of the beam was varied between $\phi = 0^{\circ}$ (normal incidence) and $\phi = 70^{\circ}$ (grazing incidence) to probe
the out-of-plane and in-plane XMCD. The XAS signal is taken as the average intensity $(\sigma^{+}+\sigma^{-})/2$ and to the XMCD as
$(\sigma^{+}-\sigma^{-})$. Hysteresis loops were measured recording the XMCD/XAS at the $L_{3}$ edge of Cobalt as a function of
magnetic field. The experimental timescales for measuring spectra as well as ramping of the magnets to their designated values are
of the order of 10-100s. During the experiments no changes in the Xenon lines induced by the heat load of the x-rays were detected.

\end{document}